\documentclass[useAMS,usenatbib]{mn2e}
\usepackage{natbib, graphicx, times}
\usepackage{amsmath}
\usepackage{epstopdf}
\usepackage{amssymb}
\usepackage{extarrows}
\usepackage{color}
\usepackage{multirow}
\newcommand{\apj}{ApJ}
\newcommand{\apjl}{ApJ}
\newcommand{\apjs}{ApJS}
\newcommand{\aap}{A \& A}
\newcommand{\araa}{ARA\&A}
\newcommand{\aj}{AJ}
\newcommand{\mnras}{MNRAS}

\newcommand{\nat}{Nature}

\newcommand{\out}{_\mathrm{out}}
\newcommand{\oct}{_\mathrm{Oct}}
\newcommand{\mig}{_\mathrm{mig}}
\newcommand{\dis}{_\mathrm{dis}}

\renewcommand{\lim}{_\mathrm{lim}}
\renewcommand{\max}{_\mathrm{max}}
\renewcommand{\min}{_\mathrm{min}}

\newcommand{\equ}[1]{Equation~(\ref{#1})}
\newcommand{\equp}[1]{(Equation~\ref{#1})}
\newcommand{\equnone}[1]{Equation~\ref{#1}}

\def\cFHJ{{\mathcal{F}_\mathrm{HJ}^\mathrm{LK}}}
\def\cFdis{{\mathcal{F}_\mathrm{dis}^\mathrm{LK}}}
\def\cFmig{{\mathcal{F}_\mathrm{mig}^\mathrm{LK}}}

\newcommand{\HJ}{_\mathrm{HJ}}

\title[Lidov-Kozai migration fractions]{The formation efficiency of
  close-in planets via Lidov-Kozai migration: analytic calculations}
\author[Mu\~noz, Lai $\&$ Liu]{Diego J. Mu\~noz$^{1}$\thanks{E-mail:dmunoz@astro.cornell.edu},
Dong Lai$^{1}$
and Bin Liu$^{2}$\\
$^{1}$ Cornell Center for Astrophysics and Planetary Science, 
Department of Astronomy, Cornell University, Ithaca, NY 14853, USA\\
$^{2}$ Center for Astrophysics, University of Science and Technology of China, Hefei, Anhui 230026, People's Republic of China
}

\begin{document}


\pagerange{\pageref{firstpage}--\pageref{lastpage}} \pubyear{2016}

\maketitle

\label{firstpage}

\begin{abstract}
Lidov-Kozai oscillations of planets in stellar binaries, combined with
tidal dissipation, can lead to the formation of hot Jupiters (HJs) or
tidal disruption of planets. Recent population synthesis studies
have found
that the fraction of systems resulting in HJs (${\cal F}_{\rm HJ}$)
depends strongly on the planet mass, host stellar type and tidal
dissipation strength, while the total migration fraction ${\cal
  F}_{\rm mig} ={\cal F}_{\rm HJ}+{\cal F}_{\rm dis}$ (including both
HJ formation and tidal disruption) exhibits much weaker dependence.
We present an analytical method for calculating ${\cal F}_{\rm HJ}$ and
${\cal F}_{\rm mig}$ in the Lidov-Kozai migration scenario. The key
ingredient of our method is to determine the critical initial
planet-binary inclination angle that drives the planet to reach
sufficiently large eccentricity for efficient tidal dissipation or
disruption. This calculation includes the effects of octupole potential and
short-range forces on the planet. Our analytical method reproduces the resulting planet 
migration/disruption fractions from population synthesis, and can be easily 
implemented for various planet, stellar/companion types, and for different distributions 
of initial planetary semi-major axes, binary separations and eccentricities. We extend 
our calculations to planets in the super-Earth mass range and discuss the conditions for
 such planets to survive Lidov-Kozai migration and form close-in rocky planets.
\end{abstract}


\section{Introduction}

The occurrence of ``hot Jupiters" (HJs) -- gas giants
with periods $\lesssim 5$~days
-- is estimated to be around $1\%$ for FGK stars
\citep{mar05,wri12,how12,fre13,pet13}.  It is 
commonly believed that these planets formed at larger separations from their host stars
(presumably beyond the ice line) and subsequently
``migrated" to semi-major axes of less than $0.1$~AU. The known
migration mechanisms can be (i) ``disc mediated" 
due to planet-disc interaction \citep[e.g.,][]{gol80,lin96}, 
(ii) ``planet mediated", 
including strong planet-planet scatterings 
and various forms of secular interactions among multiple planets 
\citep[e.g.][]{ras96,cha08,jur08,nag08,wu11,bea12,pet15b},
or (iii) ``binary mediated", i.e., secular interactions with a distant stellar companion
\citep{wu03,wu07,fab07,nao12,cor11,sto14,pet15a,and16}.
Excluding disc-driven
migration, these mechanisms require the migrating planet to reach high
eccentricities in order for tidal dissipation at pericenter 
to shrink the planet semi-major axis down to $\lesssim 0.1$~AU. Thus, such mechanisms are
sometimes referred to collectively as ``high-$e$ migration".

Planet migration following eccentricity excitation through Lidov-Kozai oscillations
\citep{lid62,koz62} induced by a stellar companion\footnote{LK oscillations induced by a planetary companion can play an important role 
in some planet-mediated high-$e$ migration scenarios. In this paper, we focus on
LK oscillations induced by a stellar-mass companion.}
-- often called Lidov-Kozai (LK) migration --
has been investigated by different researchers via Monte Carlo experiments
\citep{wu07,fab07,nao12,pet15a,and16}.
The latest studies have concluded that at most $20\%$
of observed HJs can be explained by LK migration in stellar binaries. 
Similarly, observational evidence \citep{daw13,daw15}
also indicates that the population of HJs is unlikely to be
explained by a single migration mechanism.

The occurrence rate of HJs produced by LK migration in stellar binaries can be computed via
\begin{equation}\label{eq:occurrence_rate}
\mathcal{R}\HJ^\mathrm{LK} = \mathcal{F}_b \times \mathcal{F}_p \times \mathcal{F}_\mathrm{HJ}^\mathrm{LK},
\end{equation}
where $\mathcal{F}_b$ is fraction of stars having a wide binary companion,
$\mathcal{F}_p$ is the fraction of solar-type stars
hosting a ``regular'' giant planet at a distance of a few AU, 
and $\mathcal{F}_\mathrm{HJ}^\mathrm{LK}$ is the formation 
fraction/efficiency of HJs
(i.e., the fraction of systems that result in HJs for a given 
distributions of binaries and ``regular'' planets).
Optimistic values of $\mathcal{F}_b{\sim}20\%$ 
(for binary separations between 100~AU and 1000~AU; see \citealp{rag10})
and $\mathcal{F}_p{\sim}15\%$ \citep{pet15a} require
$\mathcal{F}_\mathrm{HJ}^\mathrm{LK}{\sim}30\%$--$40\%$ 
to explain the observed HJ occurrence rate ($\sim 1\%$) using LK migration alone.

In a recent work, \citet{and16} (hereafter ASL) 
conducted a comprehensive population synthesis study of HJ formation
from LK oscillations in stellar binaries, including all relevant
physical effects (the octupole potential from the binary companion,
mutual precession of the host stellar spin axis and planet orbital
axis, tidal dissipation, and stellar spin-down due to magnetic
braking). Unlike previous works, ASL considered a range of planet
masses ($0.3-5 M_J$) and initial semi-major axes ($a_0=1-5$~AU),
different properties for the host star, and varying tidal dissipation
strengths.  They found that the HJ formation efficiency depends
strongly on planet mass and stellar type, with $\cFHJ = 1{-}4\%$
(depending on tidal dissipation strength) for $M_p = 1\,M_J$ , and
larger (up to $8\%$) for more massive planets. The production
efficiency of ``hot Saturns'' ($M_p = 0.3\,M_J$) is much lower,
because most migrating planets are tidally disrupted. Remarkably, ASL
found that the fraction of migrating systems, i.e. those that result
in either HJ formation or tidal disruption, 
$\cFmig\equiv \cFHJ+\cFdis\simeq 11{-}14\%$
is roughly constant, having little variation with planet mass, stellar
type and tidal dissipation strength. ASL provided a qualitative explanation
of how such a robust $\cFmig$ may come about (see Sections 3.4 and 5.4.1 of that work).
Note that \citep{pet15a} considered $M_p=1\,M_J$ around solar-type stars and found a similar
$\cFHJ$ as ASL, but a much larger disruption fraction ($\cFdis\sim
25\%$), and thus obtained $\cFmig\approx 27{-}29\%$.  The difference
arises from the fact that \citep{pet15a} placed all planets
initially at $a_0=5$~AU and assumed that the binary eccentricity
distribution extends up to 0.95, whereas ASL considered $a_0=1{-}5$~AU
and a maximum binary eccentricity of 0.8 (see Section 6.2 of ASL
for further discussion).

In this paper, following on the work of ASL, we present an analytic
study that explains (i) how the LK migration fraction of close-in planets
are largely determined by geometrical considerations,
and (ii) how the planet survival and disruption fractions 
depend on the properties of the planets and host stars.
We outline the necessary steps that allow for simple calculations of
$\cFmig,~\cFHJ$ and $\cFdis$ under various conditions and assumptions (e.g.,
the distributions of initial planetary semi-major axes, binary
separations and eccentricities, the masses of planet and host star).  Our
analytic calculations agree with the results of population
synthesis studies. In addition, we extend our calculations to planets
in the super-Earth mass range.

\section{Lidov-Kozai Oscillations with Octupole Effect and Short-Range Forces: Maximum
Eccentricity}

The success of LK migration depends on the maximum eccentricity
$e\max$ that can be achieved by the planet in the LK cycles. This
eccentricity needs to be high enough such that the pericenter
separation between the planet and its host star is sufficiently small
for tidal dissipation in the planet to be efficient (see
Section~\ref{sec:required_distance}).  The maximum eccentricity
depends on the initial inclination $i_0$ between the planetary and
binary orbital axes, and on various system parameters: $a_0$ (initial planetary
semi-major axis), $a\out$ and $e\out$ (binary semi-major axis and
eccentricity), $M_p$, $M_*$, $M\out$ (masses of the planet, host star
and outer binary companion), and the planet's radius $R_p$ and its
internal structure (Love number).

In \citet{liu15} (hereafter LML), 
we have studied the maximum eccentricity in LK
oscillations including the octupole potential of the binary companion and the
effect of short-range forces (associated with General Relativity, the
tidal bulge and rotational bulge of the planet). Here we
summarize the key results needed for this paper.

The strength of the octupole potential from the binary companion relative to the
quadrupole potential is characterized by the dimensionless ratio
\begin{equation}\label{eq:octupole}
 \epsilon\oct\equiv \frac{a_0}{a\out}\frac{e\out}{1-e\out^2}~~.
\end{equation}
When $\epsilon\oct\rightarrow 0$, the octupole effect can be neglected. 
In this limit, the planet's dimensionless angular momentum (projected
along the binary axis) $j_z=\sqrt{1-e^2}\cos i$ is conserved,
resulting in regular LK oscillations in eccentricity and inclination. 
Together with the conservation of energy, the maximum eccentricity
$e\max$ can be calculated analytically as a function of $i_0$ (and other parameters) \citep{fab07}.
Assuming that the planet has zero initial eccentricity, $e\max$ is determined by (LML)
\begin{equation}\label{eq:max_ecc}
\begin{split}
&\epsilon_\mathrm{GR}\bigg(\frac{1}{j\min}-1\bigg)
+\frac{\epsilon_\mathrm{Tide}}{15}\bigg(\frac{1+3e\max^2+\frac{3}{8}e\max^4}{j\min^9}-1\bigg)\\
&+\frac{\epsilon_\mathrm{Rot}}{3}\bigg(\frac{1}{j\min^3}-1\bigg)
=\frac{9}{8}\frac{e^2\max}{j\min^2}\bigg(j\min^2-\frac{5}{3}\cos^2\!i_0\bigg)~,
\end{split}
\end{equation}
where $j\min \equiv\sqrt{1-e^2\max}$, and the dimensionless quantities
\begin{subequations}\label{eq:srf}
\begin{align}
\epsilon_\mathrm{GR}&\equiv\frac{3\mathcal{G} M_*^2a\out^3(1-e\out^2)^{3/2}}{a_0^4c^2M\out}~\\
\epsilon_\mathrm{Tide}&\equiv\frac{15M_*^2a\out^3(1-e\out^2)^{3/2}k_{2p}R_p^5}{a_0^8M_p M\out}~\\
\epsilon_\mathrm{Rot}&\equiv\frac{M_* a\out^3(1-e\out^2)^{3/2}k_{qp}\Omega_{p}^2R_p^5}{\mathcal{G}a_0^5M_pM\out}~,
\end{align}
\end{subequations}
represent the relative strengths of apsidal precessions due to General
Relativity (GR), tidal bulge, and rotational distortion of the planet.
Here $k_{2p}$ is the Love number, $k_{qp}$ the
apsidal motion constant and $\Omega_p$ the rotation rate of the
planet.  In the ``pure" LK limit ($\epsilon_\mathrm{GR}=\epsilon_\mathrm{Tide}=
\epsilon_\mathrm{Rot}=0$), \equ{eq:max_ecc} reduces to the
standard expression $e\max=\sqrt{1-(5/3)\cos^2i_0}$.  For
typical planetary rotation rates, the contribution from the rotational
bulge ($\epsilon_\mathrm{Rot}$) can be neglected compared to the tidal
term ($\epsilon_\mathrm{Tide}$) (see LML)

The maximum value of $e\max$ is achieved when $i_0=90^\circ$ in \equ{eq:max_ecc}, and is referred to
as the ``limiting eccentricity" $e\lim$. When $e\lim\approx1$, the limiting eccentricity
satisfies (Eq.~53 of LML)
\begin{equation}\label{eq:lim_ecc}
\frac{\epsilon_\mathrm{GR}}{(1-e\lim^2)^{1/2}}
+\frac{7}{24}\frac{\epsilon_\mathrm{Tide}}{(1-e\lim^2)^{9/2}}\simeq\frac{9}{8}~.
\end{equation}

\begin{figure}
\includegraphics[width=0.49\textwidth]{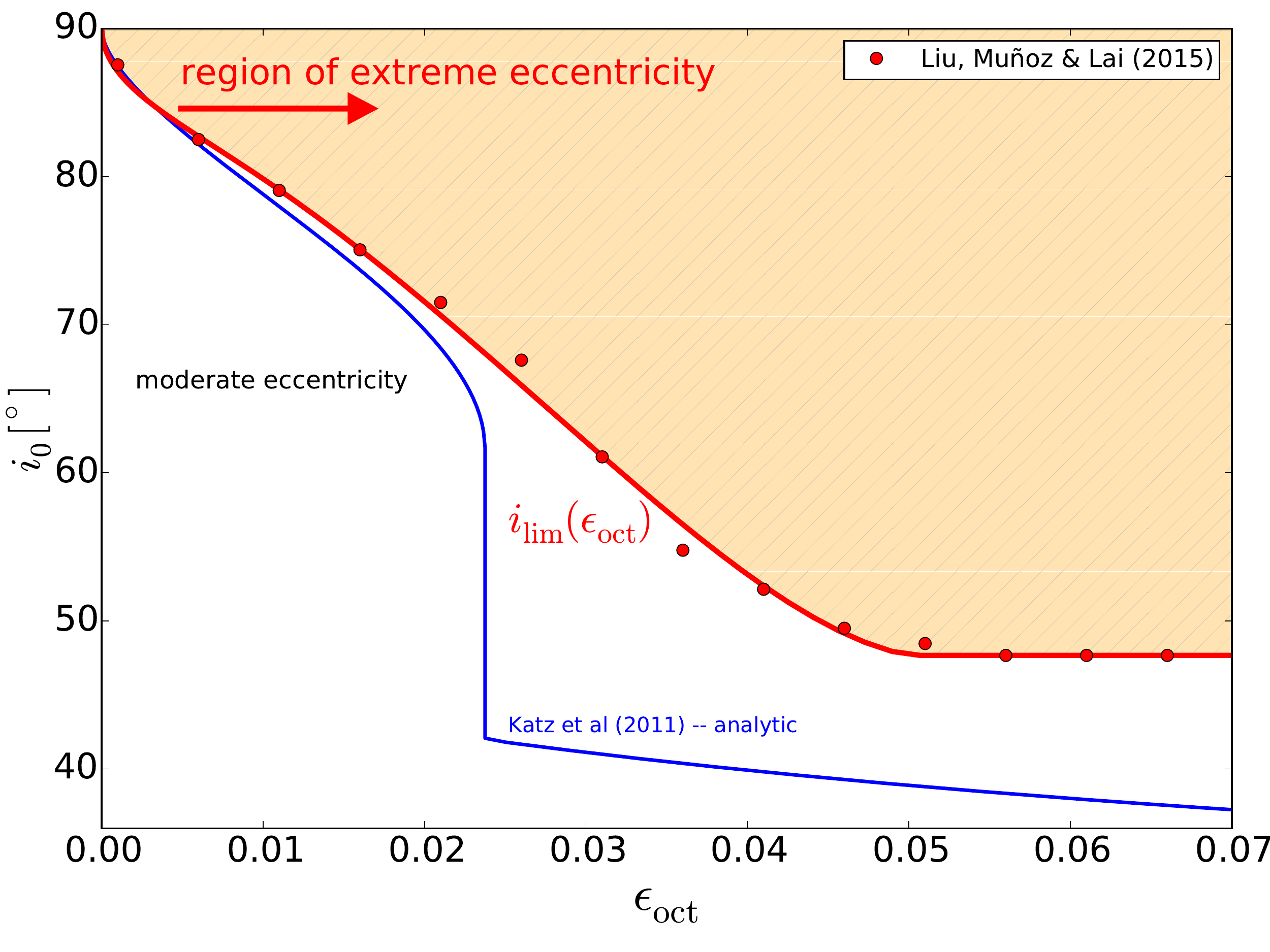}
\vspace{-0.25in}
\caption{The behavior of LK oscillations with the octupole and short-range-force (SRF) effects
  can be divided into ``moderate'' eccentricity evolution and
  ``extreme'' eccentricity evolution. The boundary between these two
  regions is a function of $\epsilon\oct$ (the dimensionless octupole strength) and
  $i_0$ (the initial inclination angle).  In the $\epsilon\oct$-$i_0$
  plane, this boundary is represented by the critical initial angle
  $i\lim(\epsilon\oct)$ above which planets can reach the limiting
  eccentricity $e\lim$ \equp{eq:lim_ecc}. The data points are taken from
  the numerical integrations of \citet{liu15} and the red curve depicts
  the fitting function \equp{eq:octupole_window}.
For $\epsilon\oct \gtrsim 0.05$, $i\lim(\epsilon\oct)$ flattens out to
a constant value $\approx 48^\circ$.  For comparison, the blue curve
shows the analytic result for the critical $i_0$ for orbital
flip without SRF, as obtained by \citet{kat11}.
\label{fig:critical_angle}}
\end{figure}

In the presence of finite octupole potential ($\epsilon_{\rm Oct}\neq 0$), $j_z$ is no longer a constant,
but a slowly varying quantity \citep{for00,lit11,kat11}, with $dj_z/dt\propto \epsilon_{\rm Oct}$. 
For sufficiently large initial inclinations ($i_0>i\lim$), 
the slow evolution of $j_z$ can take it to extremely small values and even make it cross
zero \citep[i.e. $i\rightarrow90^\circ$, see e.g.,][]{kat11}. As $j_z$ becomes small, $e$ can reach very high values.

As described by LML,
the effect of the octupole is to
increase the range of possible initial angles $i_0$ for which $e\lim$ can be reached. 
To capture the effect described by LML,
we can approximate the maximum eccentricity with
\footnote{
\equ{eq:max_ecc_oct} is restricted to $i_0\le 90^\circ$.
However, note that the system is symmetric around $90^\circ$, 
thus the limiting eccentricity is also reached  for $90^\circ<i_0<180-i\lim$.}
\begin{equation}\label{eq:max_ecc_oct}
e\max(i_0,\epsilon\oct)\simeq \left\{
\begin{array}{lc}
e_\mathrm{max}^\mathrm{(Quad)}(i_0) &  i_0 < i\lim(\epsilon\oct) \\\\
e\lim  & i_0 \geq i\lim(\epsilon\oct)
\end{array}
\right.
\end{equation}
where $e_\mathrm{max}^\mathrm{(Quad)}(i_0)$ is the solution to \equ{eq:max_ecc}
and $e\lim$ is the solution to \equ{eq:lim_ecc},
and $i\lim$ is the critical angle above which $e\lim$ is reached
(for $\epsilon\oct\rightarrow0$, it is required that $i\lim\rightarrow90^\circ$,
such that the quadrupole approximation is recovered).
Note that \equ{eq:max_ecc_oct} neglects the ``transition
region'' for $i_0$ just below $i\lim$ in which the octupole potential
makes $e_{\rm max}$ larger than $e_\mathrm{max}^\mathrm{(Quad)}$ but
not as large as $e\lim$. The numerical calculations of LML
showed that 
when tidal distortion ($\epsilon_{\rm Tide}$) dominates the SRFs, 
the transition region is narrow.

The critical inclination angle $i\lim$, which defines
the ``octupole-active" window, depends on $\epsilon\oct$.
Figure \ref{fig:critical_angle} shows the numerically computed values of
of $i\lim$ (LML).
These can be well approximated by the fitting formula
\begin{equation}\label{eq:octupole_window}
\cos^2i\lim= A\bigg(\frac{\epsilon\oct}{0.1}\bigg)
+B\bigg(\frac{\epsilon\oct}{0.1}\bigg)^2
+C\bigg(\frac{\epsilon\oct}{0.1}\bigg)^3
+D\bigg(\frac{\epsilon\oct}{0.1}\bigg)^4
\end{equation}
for $\epsilon\oct\leq0.05$ with $A=0.26$, $B=-0.536$, $C=12.05$ and  $D=-16.78$.
For $\epsilon\oct>0.05$, the dependence on $\epsilon_{\rm oct}$ flattens off at 
$\cos^2i\lim=0.45$ or $i\lim=48^\circ$ (Fig.~\ref{fig:critical_angle}).
Note that, as shown by LML,
the octupole-active' window is equivalent to the 
range of inclinations susceptible to orbital ``flips" in LK cycles
{\it without } SRFs.  \citet{kat11} has obtained an
analytic expression of the critical inclination angle for orbital flip 
-- this analytic result is valid only for $\epsilon_{\rm oct}\ll 1$
(see Fig.~\ref{fig:critical_angle}).

We note that, in general, the octupole-active window of extreme eccentricity is
reached only when $\omega$ (the argument of pericenter) of the planet
is circulating \citep[as discussed by][]{kat11}.
However, since planets commonly start with $e\simeq 0$, this distinction
is irrelevant, as planets at $e\simeq 0$ 
lie very close to the separatrix between the librating and
circulating regions of phase space.

\section{Derivation of migration fractions}
%

\subsection{Required pericenter distance for migration or disruption}\label{sec:required_distance}

In order to migrate, a distant planet must reach small pericenter distance,
$r_p=a_0 (1-e)$, 
such that tidal dissipation can 
reduce the planet's semi-major axis within a few Gyr. 
The effective orbital decay rate during LK oscillations can be estimated from
the orbital decay rate due to tidal dissipation at $e=e_{\rm max}$ multiplied by the fraction of the time the 
planet spends in the high-$e$ phase ($e\sim e\max$), giving (see Section 3.2 and Eq.~32
of ASL)
\begin{equation}
\begin{split}
\frac{1}{t_\mathrm{dec}}&\equiv\left|\frac{1}{a}\frac{da}{dt}\right|_\mathrm{Tide,LK}\\
&\simeq 6 k_{2p} \Delta t_L \frac{M_*}{M_p}\left(\frac{R_p}{a_0}\right)^5n^2
\frac{f_1(e_\mathrm{max})}{j_\mathrm{min}^{14}}\\
&\simeq 2.8 \, k_{2p} \Delta t_L \frac{\mathcal{G}M_*^2}{M_p}\frac{R_p^5}{a_0~r_{p,\mathrm{min}}^{7}} 
\end{split}
\end{equation}
where ${\Delta}t$ is the lag time, the pericenter distance at maximum eccentricity is
defined as $r_{p,\mathrm{min}}\equiv a_0(1-e_\mathrm{max})$ and
the expression $f_1(e_\mathrm{max})$ is replaced by its approximate value 
$3861/64$.
In order to have migration within $\sim 10^9$~years (i.e., $t_{\rm dec}\lesssim 10^9$~years), we
require $r_{p,\mathrm{min}}\lesssim r_{p,\mathrm{mig}}$, with
\begin{equation}\label{eq:mig_distance}
\begin{split}
r_{p,\mathrm{mig}}
&\equiv2.4\times10^{-2}~\bigg(\frac{R_p}{R_\mathrm{J}}\bigg)^{\tfrac{5}{7}}~\text{AU}\\
&\times\bigg(\frac{\chi}{10}\bigg)^{1/7}
\bigg(\frac{M_p}{M_\mathrm{J}}\bigg)^{-\tfrac{1}{7}}
 \bigg(\frac{a_0}{1~\text{AU}}\bigg)^{-\tfrac{1}{7}}
\bigg(\frac{k_{2p}}{0.37}\bigg)^{\tfrac{1}{7}}
\bigg(\frac{M_*}{1M_\odot}\bigg)^{\tfrac{2}{7}}~,
\end{split}
\end{equation}
where $\chi\equiv{\Delta}t/(0.1{\rm s})$ is a dimensionless tidal enhancement factor.

We see that, 
except for planet radius, the dependence of the critical pericenter
distance $r_{p,{\rm mig}}$ on the physical parameters of the system is weak, and thus it
represents a very robust (albeit not an exact) estimate of how close
planets must get to their host stars in order to migrate effectively (see ASL).
For this pericenter to be reached, a planet initially located at $a_0$ needs to achieve a maximum
eccentricity of
\begin{equation}\label{eq:mig_eccentricity}
e\mig\equiv1 -\frac{r_{p,{\rm mig}}}{a_0}
\end{equation}
via LK oscillations.
Note that,
since the final semi-major axis $a_F$ of the planet after tidal decay is approximately
$a_F\simeq r_{p,\mathrm{mig}}(1+e\max)\simeq
2r_{p,\mathrm{mig}}$, \equ{eq:mig_distance} indicates that 
it is difficult to form HJs with $a_F\gtrsim 0.06$~AU via LK migration (\citealp{pet15a};
ASL), unless the tidal dissipation parameter $\chi$ is much larger than the canonical value of 10.

Equation (\ref{eq:mig_distance}) sets the maximum pericenter separation that allows
for LK migration. There is also a {\it minimum} pericenter separation that permits 
``successfully migrated'' planets.
For pericenter separations smaller than the tidal radius
\begin{equation}\label{eq:dis_distance}
\begin{split}
r_{p,\mathrm{dis}}&=\eta R_p\left(\frac{M_*}{M_p}\right)^{1/3}\\
&= 1.3\times10^{-2}~\text{AU} \left(\frac{R_p}{R_\mathrm{J}}\right)\left(\frac{M_p}{M_\mathrm{J}}\right)^{-1/3}\left(\frac{M_*}{M_\odot}\right)^{1/3}~~.
\end{split}
\end{equation}
\citep[where $\eta=2.7$;][]{gui11}, planets will be disrupted and therefore unable to turn into HJs.
As before,
for this specific pericenter to be reached, a planet initially located at $a_0$ needs to achieve LK maximum
eccentricity of
\begin{equation}\label{eq:dis_eccentricity}
e_\mathrm{dis} \equiv1 -\frac{r_{p,\mathrm{dis}}}{a_0}~~.
\end{equation}
Note that, if $r_{p,\mathrm{dis}}>r_{p,\mathrm{mig}}$, 
all migration-inducing LK oscillations will result in tidal disruptions. This condition imposes
a lower limit in the planet mass that permits the formation of close-in planets:
\begin{equation}\label{eq:dis_eccentricity}
M_p>0.04~M_\mathrm{J}
\bigg(\frac{R_p}{R_\mathrm{J}}\bigg)^{\tfrac{3}{2}}
\bigg(\frac{\chi}{10}\bigg)^{-\tfrac{3}{4}}
\bigg(\frac{M_*}{1M_\odot}\bigg)^{\tfrac{1}{4}}
 \bigg(\frac{a_0}{1~\text{AU}}\bigg)^{\tfrac{3}{4}}
\bigg(\frac{k_{2p}}{0.37}\bigg)^{-\tfrac{3}{4}}~~~.
\end{equation}
This implies that Saturn-mass planets 
can rarely migrate successfully via LK oscillations without first being disrupted (see ASL).

\subsection{Migration and disruption fractions}
Having defined the eccentricities required for migration and tidal disruption,
we can now compute the fractions of migration and tidal disruption for given 
$a_0$, $a\out$ and $e\out$ (plus the other fixed parameters of the systems such as
$M_p$, $M_*$, $M\out$, $R_p$ and $\chi$). To achieve $e_{\rm mig}$, the initial planetary
inclination angle (relative to the binary) must be sufficiently large, i.e., 
$i_{0,{\rm mig}}<i_0<180^\circ-i_{0,{\rm mig}}$. 
For an isotropic distribution of the initial planetary angular
momentum vectors, those that can lead to $e_{\rm mig}$ or larger 
occupy a solid angle $4\pi\cos i_{0,\mathrm{mig}}$.  Thus the migration fraction is 
\begin{equation}\label{eq:mig_fraction}
f_\mathrm{mig}^\mathrm{LK}(a_0,a\out,e\out)
=\cos i_{0,\mathrm{mig}}.
\end{equation}
Similarly, the disruption fraction is given by 
\begin{equation}\label{eq:dis_fraction}
f_\mathrm{dis}^\mathrm{LK}(a_0,a\out,e\out)
=\cos i_{0,\mathrm{dis}},
\end{equation}
where $i_{0,\mathrm{dis}}$ is the critical inclination that leads to $e\max=e\dis$.
The fraction of HJ formation (migration without disruption) is simply 
\begin{equation}\label{eq:HJ_fraction}
f_\mathrm{HJ}^\mathrm{LK}(a_0,a\out,e\out)
=f_\mathrm{mig}^\mathrm{LK}-f_\mathrm{dis}^\mathrm{LK}
=\cos i_{0,\mathrm{mig}}-\cos i_{0,\mathrm{dis}}.
\end{equation}

\subsubsection{Migration/disruption fraction in quadrupole-order theory} 

For $e\mig>e\lim$, no migration is possible and we have $\cos i_{0,{\rm mig}}=0$.
For $e\mig <e\lim$, to quadrupole-order, the critical angle for migration $i_{0,\mathrm{mig}}$ 
can be computed analytically by directly solving for $\cos i_0$ in \equ{eq:max_ecc}
with $e\max=e\mig$. Thus
\begin{equation}\label{eq:solid_angle}
\cos^2 i_{0,\mathrm{mig}}^{\rm (Quad)}=\left\{
\begin{array}{lc}
C(e\mig)  & \text{if}~~~e\mig<e\lim \\
0  & \text{if}~~~e\mig>e\lim \\
\end{array}
\right.
\end{equation}
where
\begin{equation}\label{eq:solid_angle2}
\begin{split}
C(e\mig)=
\frac{3}{5}j\mig^2&-\frac{8}{15}\frac{\epsilon_\mathrm{GR}}{e\mig^2}\bigg(j\mig-j\mig^2\bigg)\\
&-\frac{8}{225}\frac{\epsilon_\mathrm{Tide}}{e\mig^2}\bigg(\frac{1+3e\mig^2+\frac{3}{8}e\mig^4}{j\mig^2}-j\mig^2\bigg)
\end{split}
\end{equation}
with $j\mig=\sqrt{1-e\mig^2}$ and in which the rotational distortion effect in \equ{eq:max_ecc} 
has been ignored for consistency. Note that the condition
$e\mig<e\lim$ is equivalent to $C(e\mig)>0$. Similarly, the critical disruption angle
$i_{0,{\rm dis}}^{\rm (Quad)}$ is given by \equ{eq:solid_angle} with $e\dis$ replacing $e\mig$.

\begin{figure*}
\includegraphics[width=0.97\textwidth]{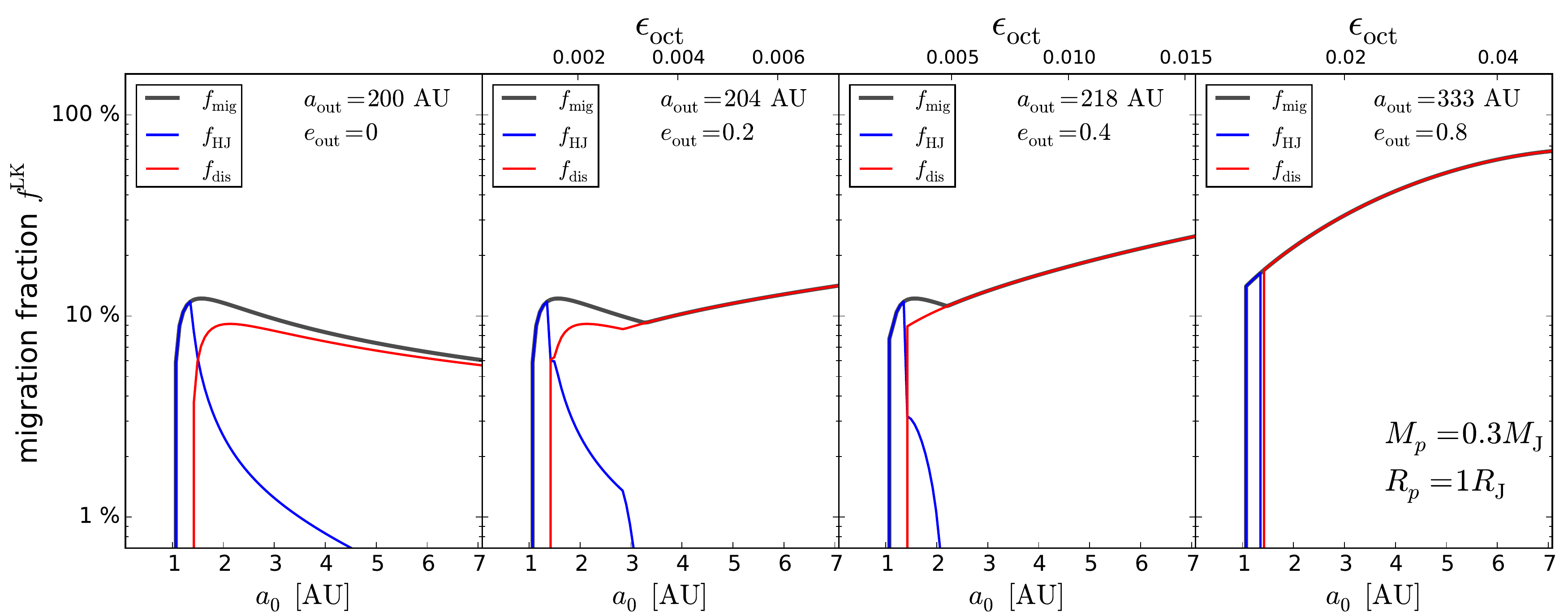}
\caption{Migration fraction $f\mig^\mathrm{LK}$
  (\equnone{eq:mig_fraction}, in black), tidal disruption fraction
  $f\dis^\mathrm{LK}$ (in red) and successful close-in planet
  formation fraction
  $f_\mathrm{HJ}^\mathrm{LK}=f\mig^\mathrm{LK}-f\dis^\mathrm{LK}$ (in
  blue) for a gas giant of mass $0.3M_\mathrm{J}$ and tidal
  dissipation parameter $\chi=10$. The first panel shows the fractions for
  a binary stellar companion with $a\out=200$~AU and $e\out=0$, for
  which the octupole terms in the binary potential
are identically zero. The second to fourth panels show increments in $e\out$ (0.2, 0.4
  and 0.8) while keeping $a\out(1-e\out^2)^{1/2}$ constant, implying a
  monotonic increase of the octupole strength $\epsilon\oct$.
If the effects of octupole potential were neglected, 
all four panels would be identical.
Upper horizontal axis shows $\epsilon\oct$,
which is directly proportional to $a_0$ for fixed $a\out$ and $e\out$. The main effect of
the octupole potential is to increase the tidal disruption fraction.
The area under the blue curve is, from left to right,
$10\%$, $7.8\%$, $5.1\%$ and $5.3\%$.
\label{fig:example_fractions}}
\end{figure*}

In the first panel of Fig.~\ref{fig:example_fractions}, we show $f_\mathrm{mig}^\mathrm{LK}$, 
$f_\mathrm{dis}^\mathrm{LK}$ and $f_\mathrm{HJ}^\mathrm{LK}$
for a $0.3M_\mathrm{J}$ planet started at different initial
semi-major axis $a_0$ and a binary companion with $a\out=200$~AU and $e\out=0$. 
We see that, for large $a_0$, SRFs are unimportant for determining $e\max$ around $e\mig$, 
thus the shape of the curve is simply $f_\mathrm{mig}^\mathrm{LK}=\cos
i_{0,{\rm mig}}=\sqrt{3/5(1-e\mig^2)}\approx\sqrt{(6/5)(r_{p,}{\mig}/a_0)}$.
For small $a_0$, $e\lim$ is smaller than the required $e\mig$, making
the migration fraction rapidly go to zero.

\subsubsection{Including the octupole effect}

When the octupole effect is included ($\epsilon\oct\neq 0$), the maximum eccentricity
achieved in LK cycles can be approximated by \equ{eq:max_ecc_oct}.
For $e\mig>e\lim$, no migration is possible and we still have
$\cos i_{0,{\rm mig}}=0$.
For $e\mig < e\lim$, since a non-negligible $\epsilon\oct$ can expand
the range of inclination angles that produce $e\mig$,
the critical angle $i_{0,\mathrm{mig}}$ 
is the lowest possible angle that results in $e\mig$, 
i.e.,  $i_{0,\mathrm{mig}}{=}{\rm min}(i_{0,\mathrm{mig}}^{\rm (Quad)},i\lim)$.
Thus, 
\begin{equation}\label{eq:solid_angle2}
\cos^2 i_{0,\mathrm{mig}}{=}{\Bigg\{}
{\begin{array}{ll}
{\mathrm{max}\bigg[C(e\mig),\cos^2 i\lim(\epsilon\oct)\bigg]}~, &{\rm if}~e\mig{\leq}e\lim \\
{0}~~,  &{\rm if}~e\mig{>}e\lim \\
\end{array}}~~~~
\end{equation}
where $C(e\mig)$ is given by \equ{eq:solid_angle2} and $\cos^2 i\lim(\epsilon\oct)$ is 
given by \equ{eq:octupole_window}. An analogous expression applies to $\cos^2i_{0,{\rm dis}}$.

We illustrate the effect of the octupole term on
$f_\mathrm{mig}^\mathrm{LK}$ and $f_\mathrm{dis}^\mathrm{LK}$ in the
three last panels of Fig.~\ref{fig:example_fractions}.  For all
panels, $a\out\sqrt{1-e\out^2}=200$~AU, while the eccentricity is
varied from 0 to 0.8.  As $e\out$ is increased, the value of
$\epsilon\oct$ for a given planet semi-major axis $a_0$ also grows.
Going from $e\out=0$ to $e\out=0.8$ illustrates the gradual transition
from $\cos^2 i_{0,\mathrm{mig}}=C(e\mig)$ for all $a_0$ (pure
quadrupole) to $\cos^2 i_{0,\mathrm{mig}}=\cos^2 i\lim(\epsilon\oct)$
for all $a_0$ (entirely dominated by the octupole effect). 
Note that the HJ fraction $f_\mathrm{HJ}^\mathrm{LK}=
f_\mathrm{mig}^\mathrm{LK}-f_\mathrm{dis}^\mathrm{LK}$ is always a
narrow distribution bounded by at the ``no-migration" limit
($e\lim<e\mig$) at small $a_0$ and by the tidal disruption limit
($e\max>e\dis$) at larger $a_0$. The width and height of this peak is
only mildly affected by the octupole contribution. On the other hand,
the fraction of tidally disrupted planets $f_\mathrm{dis}^\mathrm{LK}$
is increased significantly for larger $e\out$~. Thus, the main effect the
octupole is to increase the number of disrupted planets, rather than
the number of successfully migrated ones (\citealp{pet15a}; ASL).

\begin{figure*}
\includegraphics[width=0.97\textwidth]{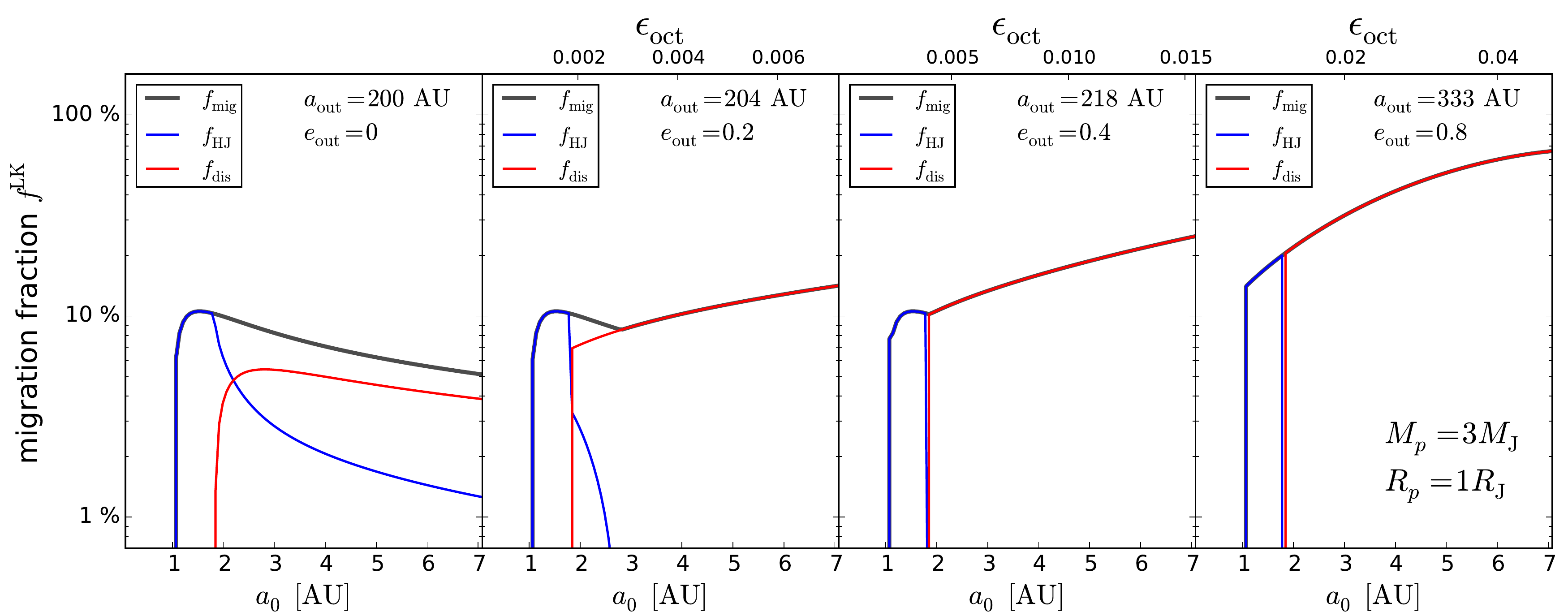}
\caption{Same as Fig.~\ref{fig:example_fractions} but for a planet ten times more massive ($M_p=3M_\mathrm{J}$).
General trends are analogous to the less massive example (the total migration curve
$f_\mathrm{mig}^\mathrm{LK}$ in black is essentially unchanged). In this case, however, planets are much more likely to survive
LK migration,  because the boundary between HJ formation and tidal disruption takes place at higher
eccentricity \equp{eq:dis_eccentricity}. The area under the blue curve is, from left to right,
$20\%$, $9.2\%$, $7.7\%$ and $13\%$.
\label{fig:example_fractions2}}
\end{figure*}

Fig.~\ref{fig:example_fractions2} shows the same examples as
Fig.~\ref{fig:example_fractions}, except for a planet mass of 
$3M_\mathrm{J}$. Although the lower limit for non-zero migration is almost
unchanged (see Eq.~47 of ASL),
the boundary between successful migration and tidal disruption 
occurs at a larger $a_0$
($e\dis$ is larger
for more massive planets, making disruption more difficult; see
\equ{eq:dis_eccentricity}.  Thus, a larger fraction of these
massive planets can survive LK oscillations and become HJs. The
mass-dependent boundary between disruption and successful migration
explains the lack of ``hot Saturns" in observations
(ASL). Note that, as in the previous example, the most
noticeable effect of the octupole terms -- which differentiate the
last three panels from the first -- is to increase the rate of tidal
disruptions, not to significantly increase the number of HJs.

\section{Total Migration Fractions as a Function of Planet Mass}
In the previous section, we described how to calculate the migration fraction
$f\mig^\mathrm{LK}$ 
and disruption fraction $f\dis^\mathrm{LK}$ 
for a given set of orbital parameters $a_0$, $a\out$ and $e\out$ 
(Eqs.~\ref{eq:mig_fraction}-\ref{eq:dis_fraction}).
We now consider the migration and disruption fractions
integrating over a range of values in $a\out$ and $e\out$: 
\begin{equation}\label{eq:integrated_fraction}
\begin{split}
{F}\mig^\mathrm{LK}(a_0)=& \int\int f\mig^\mathrm{LK}(a_0,a\out,e\out)\\
&~~~~~~~~~~\times \frac{d^2N}{d\log a\out de\out} d\log a\out de\out~~.
\end{split}
\end{equation}
For concreteness, we adopt the integration limits of $[0,0.8]$ for $e\out$
and $[2.0,3.0]$ for $\log_{10}a\out$ (ASL), although other choices are possible.
We compute this integral using the Monte Carlo method, uniformly sampling 
$e\out$  and $\log_{10}a\out$. 
\begin{figure}
\includegraphics[width=0.49\textwidth]{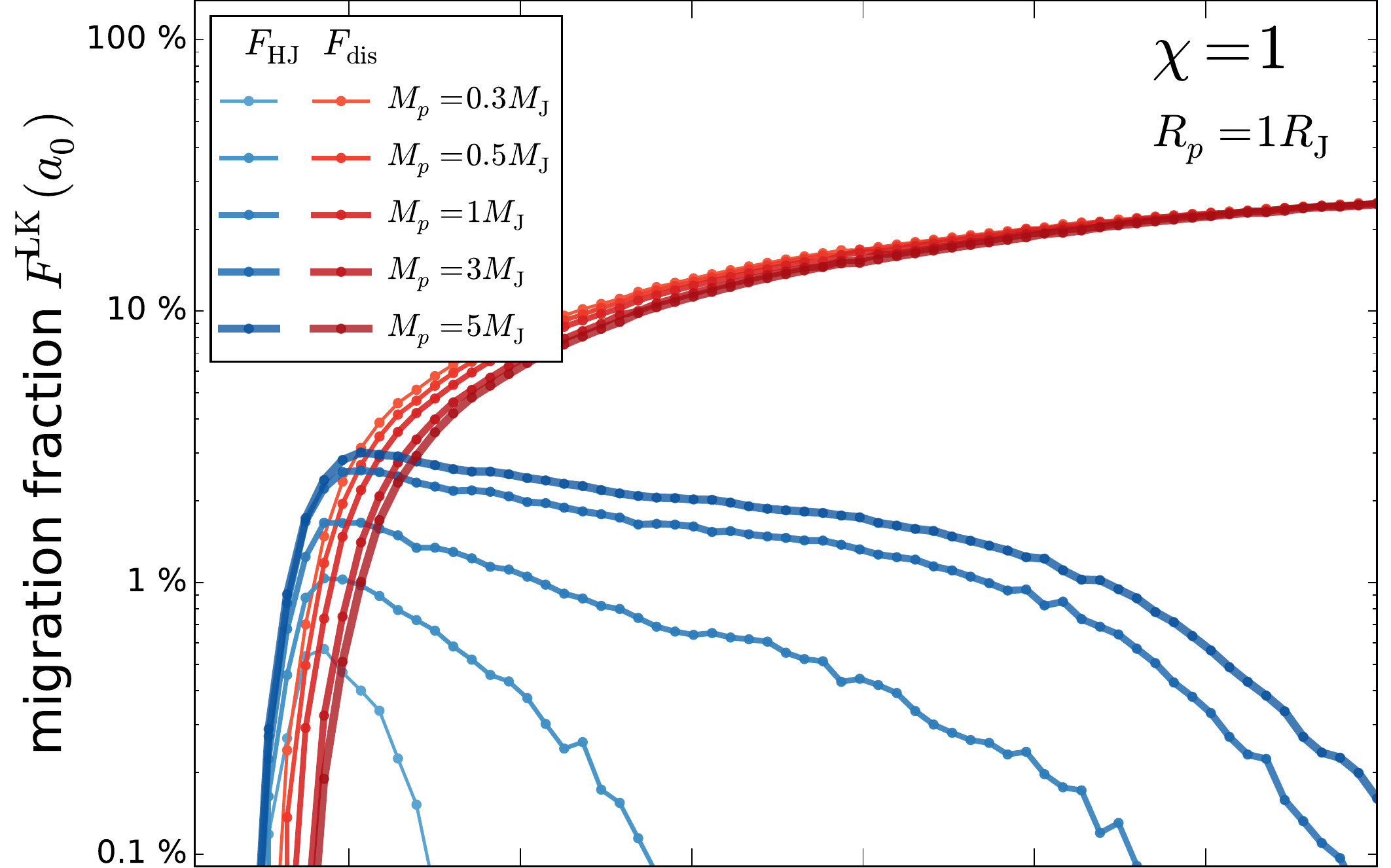}
\includegraphics[width=0.49\textwidth]{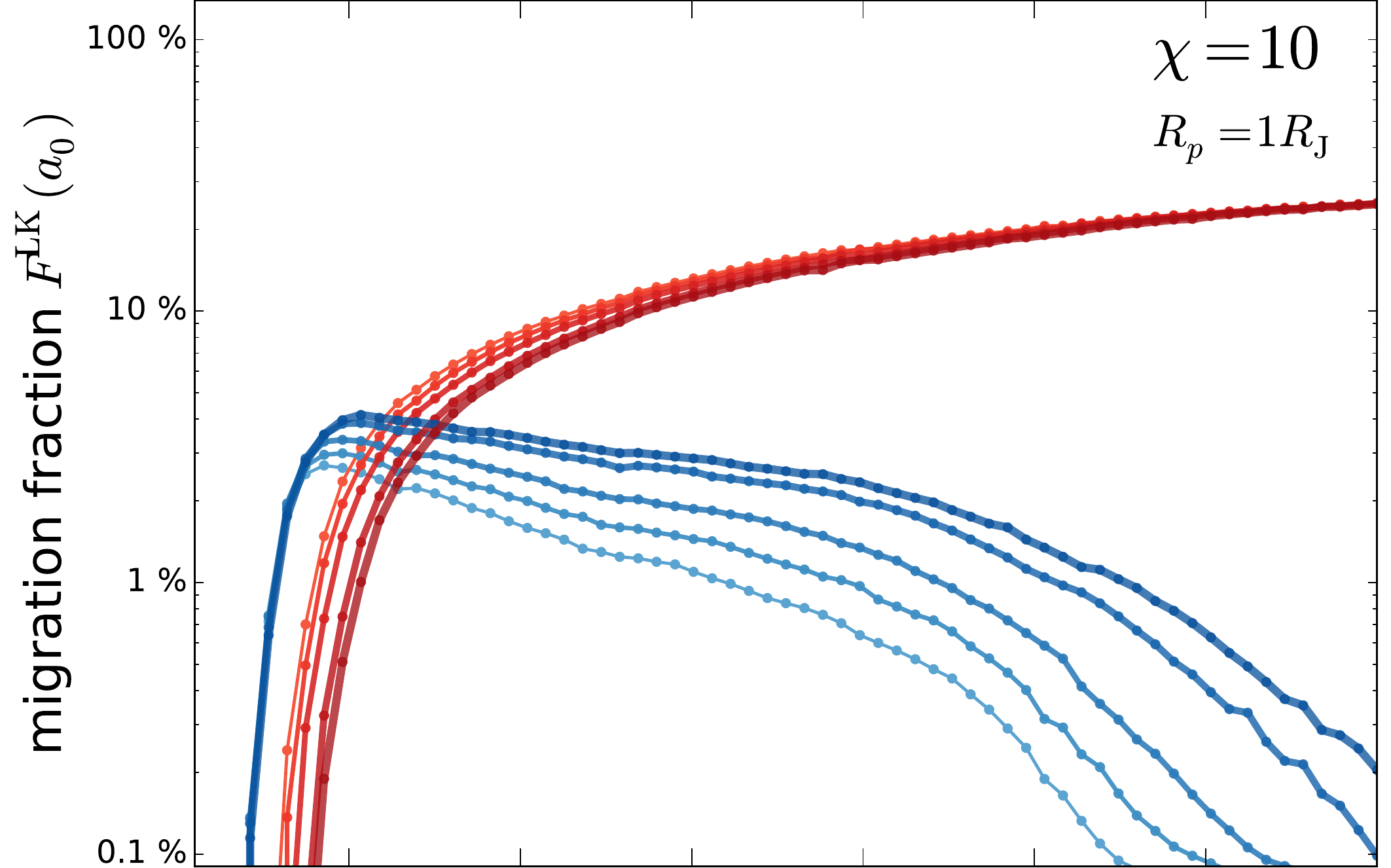}
\includegraphics[width=0.49\textwidth]{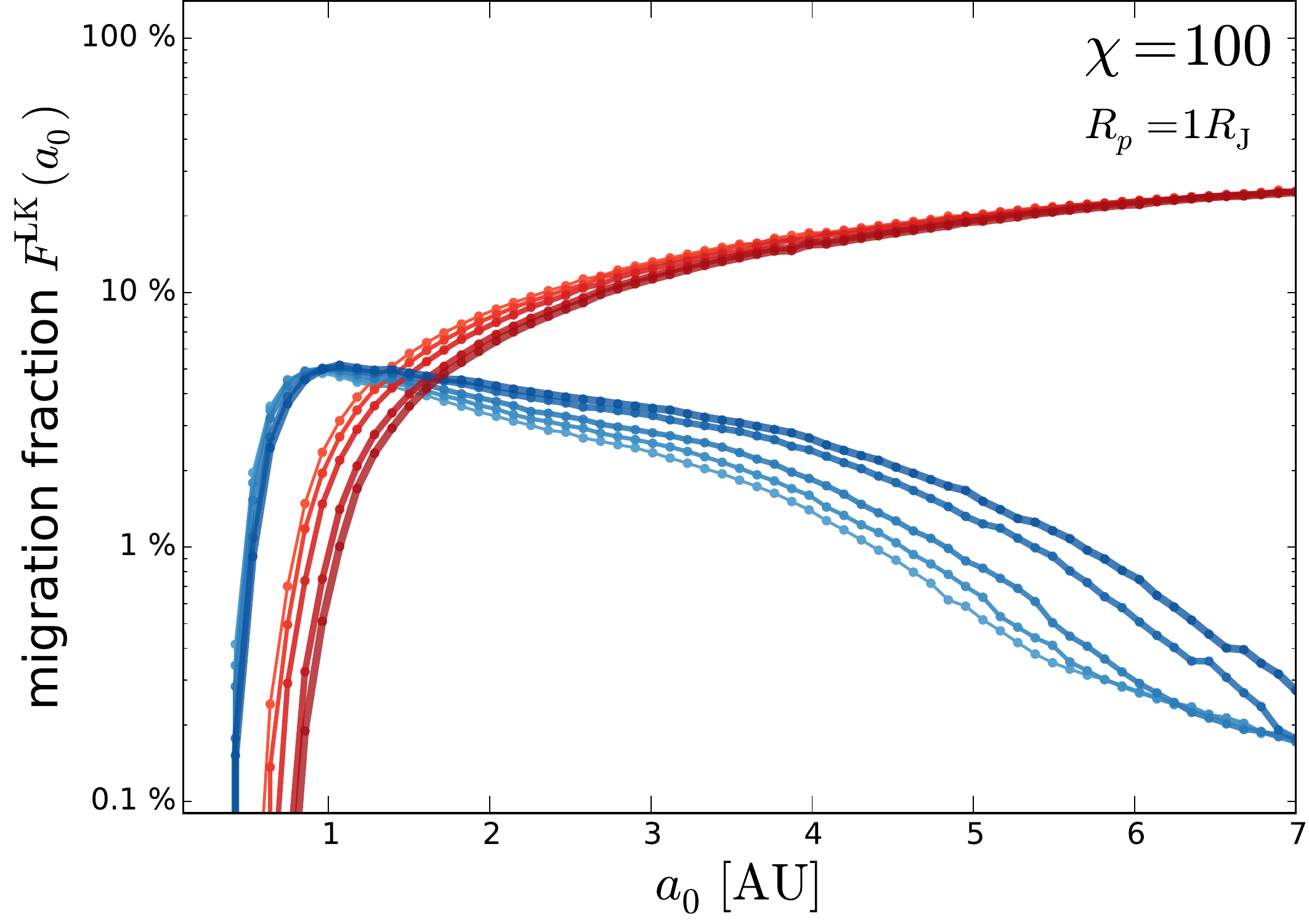}
\caption{
Integrated HJ formation fraction $F_{\rm HJ}^\mathrm{LK}(a_0)$ and
tidal disruption fraction $F_{\rm dis}^\mathrm{LK}(a_0)$ 
(see Eq.~\ref{eq:integrated_fraction})
for a range of planet masses $M_p$ and
  dimensionless tidal dissipation strengths $\chi$. Planet masses are $M_p=0.3$, 0.5, 1,
  3 and 5 $M_\mathrm{J}$ and all planet radii are taken to be
  $R_p=R_\mathrm{J}$.  
The three panels correspond to $\chi=1,\,10,\,100$.
\label{fig:integrated_fractions}}
\end{figure}
%
\subsection{Gas giants}
Fig.~\ref{fig:integrated_fractions} shows 
${F}\dis^\mathrm{LK}(a_0)$ and ${F}_{\rm HJ}^\mathrm{LK}(a_0)$ for different planet masses
and tidal dissipation strengths.
For the fiducial dissipation parameter 
($\chi=10$), the HJ fraction of Jupiter-mass objects 
is $1{-}3\%$ 
(for $a_0=1{-}5$~AU). For Saturn-mass objects, this fraction becomes
is much smaller for a wide range of $a_0$.
For smaller $\chi$, the ``hot planet'' fraction of
 Saturn-mass objects is 
$\ll 1\%$,
while only planets with
 masses above $3M_\mathrm{J}$ can successfully turn into hot planets.
 It is only in the extreme dissipation regime ($\chi=100$) that all
 gas giants across the explored mass range can turn into hot planets at rates
 of a few percent. This behavior can be easily understood from
 the definitions of $r_{p,\mig}$ and $r_{p,\dis}$
(Eqs.~\ref{eq:mig_distance} and \ref{eq:dis_distance}).
For a given $\chi$, varying the planet mass by a
 factor of a few does not change significantly the total migration
 fraction ($F\mig^\mathrm{LK}+F\dis^\mathrm{LK}$) because $r_{p,\mig}$
 is essentially unchanged. However, the relative importance of
 $F\mig^\mathrm{LK}$ and $F\dis^\mathrm{LK}$ is changed as
 $r_{p,\mathrm{dis}}$ depends on $M_p$.  
Similarly, changing $\chi$ while keeping $M_p$ fixed does not change $F\dis^\mathrm{LK}$, but
 only $F\mig^\mathrm{LK}$, as $r_{p,\mathrm{mig}}$ grows slightly with stronger
 dissipation.

These trends are in agreement with the extensive Monte Carlo
experiments of ASL.  In that work, the computed total migration
fractions (HJs plus tidal disruptions) depend very weakly on
$\chi$.  In order to compare directly with the results of ASL, we
integrate $F\mig^\mathrm{LK}(a_0)$ over a range of planet semi-major
axes. Assuming that $a_0$ is distributed uniformly between
$a_{0,\text{inner}}$ and $a_{0,\text{outer}}$, we have:
\begin{equation}\label{eq:global_integrated_fraction}
\mathcal{F}\mig^\mathrm{LK}=\int\limits_{a_{0,\text{inner}}}^{a_{0,\text{outer}}} F\mig^\mathrm{LK}(a_0) \frac{dN}{da_0} da_0~~.
\end{equation}
Using the values $a_{0,\text{inner}}=1$~AU and $a_{0,\text{outer}}=5$~AU, we compute
the integrated migration fractions $\mathcal{F}\mig^\mathrm{LK}$, $\mathcal{F}\dis^\mathrm{LK}$
and $\mathcal{F}_\mathrm{HJ}^\mathrm{LK}$ for three planet masses ($M_p=$0.3, 1.0 and  3 $M_\mathrm{J}$)
and three dissipation parameters ($\chi=1$, 10 and 100). The results 
(see Table~\ref{tab:integrated_fractions}) agree with the findings of ASL. In particular,
the total migration fraction is confirmed to be very robust and to lie in the range 
between $12\%$ and $15\%$.

\begin{table}
	\centering
	\caption{The HJ formation fraction
          $\mathcal{F}_\mathrm{HJ}^\mathrm{LK}$, tidal disruption fraction
          $\mathcal{F}\dis^\mathrm{LK}$ and total migration fraction
          $\mathcal{F}\mig^\mathrm{LK}$ for gas giants with three
          different masses and three different dissipation strengths.}
	\label{tab:integrated_fractions}
	\tabcolsep=0.07cm
	\begin{tabular}{lccccccccccc} 
		\hline
		 & &  $\chi=1$ & & & & $\chi=10$ & && & $\chi=100$  & \smallskip\\ 
		\cline{2-4} \cline{6-8} \cline{10-12} \\
		$M_p\;(M_\mathrm{J})$ &$0.3$ &  $1$ & $3$ & & $0.3$& $1$ & $3$&& $0.3$ & $1$  & $3$\\
		\cline{2-4} \cline{6-8} \cline{10-12} \\
		$\mathcal{F}_\mathrm{HJ}^\mathrm{LK}\;(\%)$& 0 & 0.8 & 1.8 &  & 1.2 & 2.1 & 2.8 &  & 2.6 & 3.1 & 3.6  \smallskip\\
		$\mathcal{F}_\mathrm{dis}^\mathrm{LK}\;(\%)$  & 12.4 & 11.6 & 10.8 &  & 12.4 & 11.6 & 10.8 &  & 12.4 & 11.6 & 10.8 \smallskip\\
		\hline
		$\mathcal{F}\mig^\mathrm{LK}\;(\%)$  & 12.4 & 12.4 & 12.6 & & 13.6 & 13.7 & 13.6 & & 15 & 14.7 & 14.4\\
		\hline
	\end{tabular}
\end{table}

\subsubsection{Other studies}

Despite the robustness of the HJ migration fraction found by ASL, this
number is not in perfect agreement with the results of other recent
works, such as \citet{nao12} and \citet{pet15a} (both considered only
$M_p=1~M_J$). The main difference with \citet{nao12} is in the choice
of $\eta$ in \equ{eq:dis_distance}. On the other hand, \citet{pet15a}
used $\eta=2.7$, the same as ASL. However, these two last works differ
in the properties of the assumed underlying population of companions
(see Section 6.2 of ASL for a discussion).  Here, we (and ASL) have
assumed the binary orbits to follow uniform distributions in $e\out$
and $\log a\out$,
with $e\out\in [0,0.8]$ and $\log a\out\in [2,3]$. Following one of
the examples of \citet{pet15a}, we choose $e\out\in [0,0.95]$ and
$\log a\out\in [2,3.2]$, and find
$\mathcal{F}_\mathrm{HJ}^\mathrm{LK}=1.2\%$,
$\mathcal{F}_\mathrm{dis}^\mathrm{LK}=21.2\%$ for a total of
$\mathcal{F}_\mathrm{mig}^\mathrm{LK}=22.4\%$, in agreement
with the results of Monte Carlo experiments.

\subsection{Rocky planets}
The procedure outlined in the previous section can be applied to rocky
planets by simple modification of the planet mass-radius relation,
Love number $k_{2p}$ and dimensionless tidal strength $\chi$. 
For Solar System rocky planets $\Delta t_L\approx600$s \citep{lam77,ner97}, i.e.,
$\chi=6000$, and $k_{2p}\approx0.3$ \citep{yod95}. We take the generic
mass-radius relation from \citet{sea07} for an ``Earth-like"
composition and calculate ${F}_\mathrm{mig}^\mathrm{LK}$ and
${F}_\mathrm{dis}^\mathrm{LK}$ as above. The successful migration rate
of ``hot planets" is, analogous to HJs,
${F}_\mathrm{hp}^\mathrm{LK}={F}_\mathrm{mig}^\mathrm{LK}-{F}_\mathrm{dis}^\mathrm{LK}$.

\begin{figure}
\includegraphics[width=0.49\textwidth]{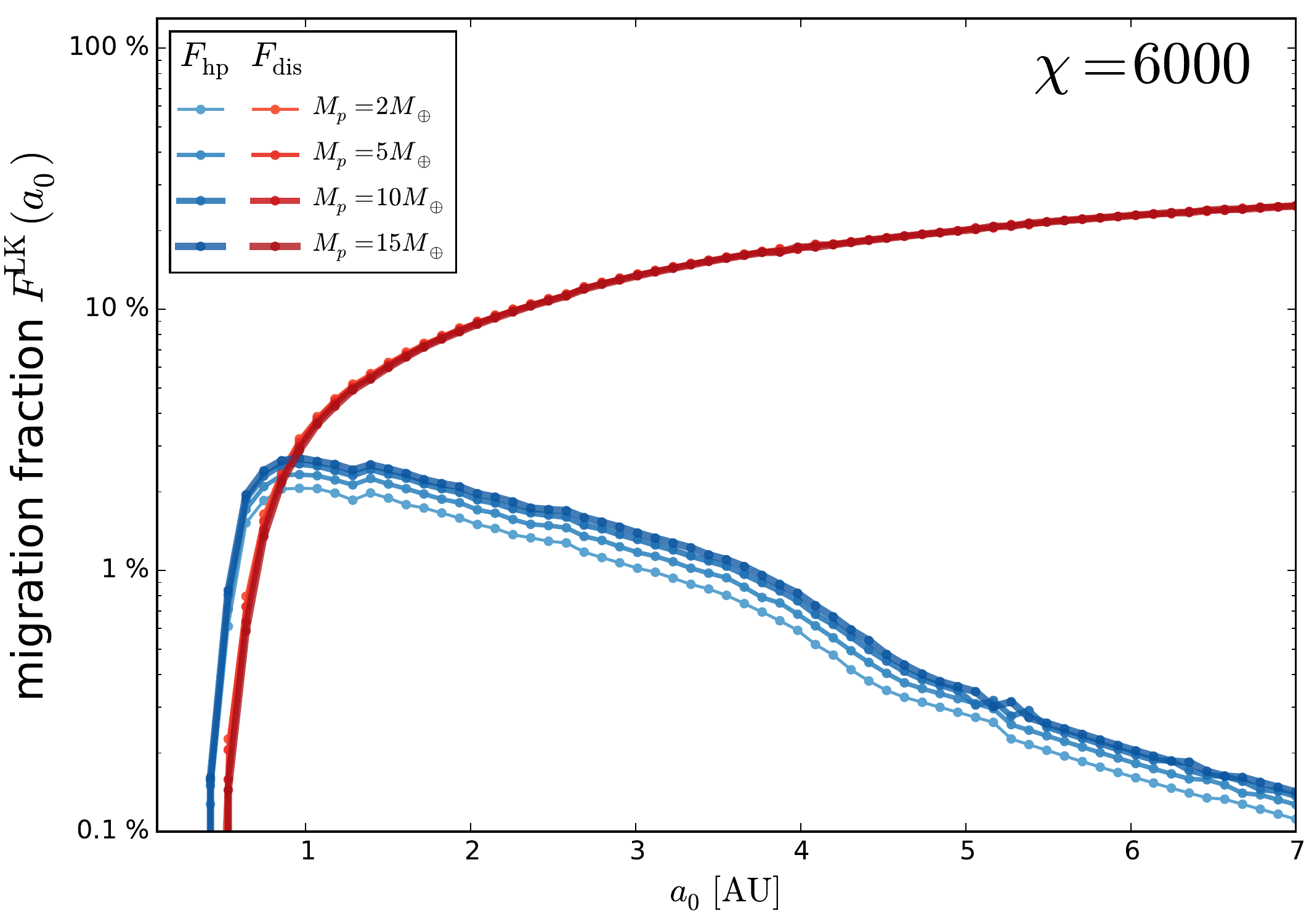}
\caption{
Integrated LK ``hot planet'' formation fraction
 $F_\mathrm{hp}^\mathrm{LK}(a_0)$ and tidal disruption fraction $F\dis^\mathrm{LK}(a_0)$ 
as in Fig.~\ref{fig:integrated_fractions} but for rocky planets. Planet
  masses are $M_p{=}2M_\oplus$, $5M_\oplus$, $10M_\oplus$ and
  $15M_\oplus$, with corresponding radii of \citep{sea07}
  $R_p{=}1.2R_\oplus$, $1.5R_\oplus$, $1.8R_\oplus$ and $2R_\oplus$.
  When integrating $F_\mathrm{hp}^\mathrm{LK}(a_0)$ and
  $F\dis^\mathrm{LK}(a_0)$ over $a_0$ between 1 and 5~AU, the global
  integrated fractions are, respectively,
  $(\mathcal{F}_\mathrm{hp}^\mathrm{LK},\mathcal{F}\dis^\mathrm{LK}){=}$
  $(1.08\%,12.9\%)$, $(1.23\%,12.9\%)$, $(1.35\%,12.8\%)$ and
  $(1.43\%,12.7\%)$. If, instead of $\chi=6000$ we use $\chi=600$, the
  hot planet fractions are reduced by half:
  $(\mathcal{F}_\mathrm{hp}^\mathrm{LK},\mathcal{F}\dis^\mathrm{LK}){=}$
  $(0.62\%,12.9\%)$, $(0.74\%,12.9\%)$, $(0.85\%,12.8\%)$ and
  $(0.92\%,12.7\%)$.
\label{fig:rocky_integrated_fractions}}
\end{figure}

In Fig.~\ref{fig:rocky_integrated_fractions}, we show
${F}_\mathrm{hp}^\mathrm{LK}$ (in blue) and
${F}_\mathrm{dis}^\mathrm{LK}$ (in red) for different rocky planets,
ranging in mass from $2M_\oplus$ to $15M_\oplus$.  The integrated
hot planet fraction $\mathcal{F}_\mathrm{hp}^\mathrm{LK}$ (for $a_0$ between 1 and 5~AU)
is between $1\%$ and $1.5\%$. Thus, these
planets are more likely to survive LK migration than Saturn-mass
planets. The reason for this is that there is enough room between
$r_{p,\mathrm{mig}}$ and $r_{p,\mathrm{dis}}$
(Eqs.~\ref{eq:mig_distance}~and~\ref{eq:dis_distance}) to
accomodate the planets into successfully migrated orbits. Rewriting
$r_{p,\mathrm{mig}}$ and $r_{p,\mathrm{dis}}$ in units more suitable
for rocky planets, we have
\begin{equation}\label{eq:mig_distance_rocky}
\begin{split}
r_{p,\mathrm{mig}}
&=2.47\times10^{-2}~\bigg(\frac{R_p}{1.2R_\oplus}\bigg)^{\tfrac{5}{7}}~\text{AU}~~~~~\\
\times&\bigg(\frac{\chi}{6000}\bigg)^{1/7}
\bigg(\frac{M_p}{2M_\oplus}\bigg)^{-\tfrac{1}{7}}
 \bigg(\frac{a_0}{1~\text{AU}}\bigg)^{-\tfrac{1}{7}}
\bigg(\frac{k_{2p}}{0.3}\bigg)^{\tfrac{1}{7}}
\bigg(\frac{M_*}{1M_\odot}\bigg)^{\tfrac{2}{7}}
\end{split} 
\end{equation}
and
\begin{equation}\label{eq:dis_distance_rocky}
r_{p,\mathrm{dis}}=7.6\times10^{-3}~\text{AU} \left(\frac{R_p}{1.2R_\oplus}\right)\left(\frac{M_p}{2M_\oplus}\right)^{-1/3}\left(\frac{M_*}{1M_\odot}\right)^{1/3}~~.
\end{equation}
Interestingly, $r_{p,\mathrm{mig}}$ is very close to the value
obtained for a Jupiter-mass planet; however, $r_{p,\mathrm{dis}}$ is
smaller, since the higher density of rocky planets pushes the Roche
limit closer to the star. Thus, provided $\chi$ is high enough (in
this case 6000), some small but non-negligible fraction of rocky
planets can survive LK oscillations and migrate into close-in orbits.
Note that, for $2M_\oplus$ rocky planets,
Eqs.~(\ref{eq:mig_distance_rocky}) and~(\ref{eq:dis_distance_rocky})
indicate that the LK migrated hot super-Earths should reside within an
orbital period range of 0.7 to 4 days.

Note that the dimensionless tidal dissipation parameter $\chi$ is highly uncertain.
Tidal dissipation in the Earth is dominated by its oceans
\citep[e.g.,][and references therein]{pea99}, which does not
necessarily apply to models of migrating super-Earths with uncertain
amounts of gas envelopes. If we assume a smaller tidal dissipation strength 
($\chi=600)$, the hot planet formation rates are reduced by a factor of two (see
the caption of Fig.~\ref{fig:rocky_integrated_fractions}). 
Such small success rates in producing close-in rocky planets are in rough agreement 
with the $N$-body experiments of \citet{pla15}, who implemented a constant-phase-lag (or
constant tidal quality factor $Q$) model to study LK migration of
small planets in the binary system $\alpha$~Cen B. In that study, only
$1{-}3$ integrations out of $\sim 600$ ended up in close-in
planets.

If we write the occurrence rate of hot rocky planets formed via LK migration as
(see \equnone{eq:occurrence_rate})
$\mathcal{R}_{\rm hp}^\mathrm{rocky}=\mathcal{F}_b\times
\mathcal{F}_p^\mathrm{rocky}\times
\mathcal{F}_\mathrm{hp}^\mathrm{LK}$, and assume that the planet
bearing fraction of rocky planets within a few AU is
$\mathcal{F}_p^\mathrm{rocky}{\sim}100\%$ \citep{pet13,fre13}, then,
with
$\mathcal{F}_b\sim 20\%$ and 
$\mathcal{F}_\mathrm{hp}^\mathrm{LK}\sim 1\%$ (see Fig.~\ref{fig:rocky_integrated_fractions}), we find
$\mathcal{R}_{\rm hp}^\mathrm{rocky}\sim0.2\%$.
Thus, an intriguing possibility is that of LK migration playing a role
in the origin of short-period rocky planets, especially for single-planet systems\footnote{
For systems containing multiple planets with a few AU semi-major axes,
LK oscillations can be suppressed by mutual planet-planet interactions if the planet mass
is too large. From Eq.~(7) of \citet{mun15b}, the critical ``shielding mass'' is 
about $3 M_\oplus$ for $a_0\sim 3$~AU and a stellar companion located at
150~AU \citep[see also][]{mar15,ham16}. For smaller planet masses,  LK oscillation 
may operate for each planet independently.}.
Observationally, ultra-short-period ($\lesssim 2$~days) rocky planets 
are found to have an occurrence rate of $\sim0.5\%$ in the {\it Kepler} catalog
\citep[e.g.][]{san14}, and the occurrence rate of rocky planets with slightly longer
periods (a few days) is appreciably higher \citep{fre13}.
While many of these are expected to be members of multiple-planet 
systems \citep{san14}, some may be true ``singles'' \citep{ste13}.
LK migration of rocky planets may contribute to this population of close-in planets.

%

\section{Summary}
We have derived a simple analytical method for calculating migration
fractions of close-in planets via Lidov-Kozai oscillations in stellar binaries.  
An important ingredient of the method is the relationship between the
maximum eccentricity $e\max$ that can be achieved in LK oscillations
as a function of the initial planet-binary inclination \equp{eq:max_ecc_oct}
-- the effects of short-range forces and octupole potential play an essential role.
Our method can be easily implemented for various planet and stellar/companion types,
and consists of the following steps:

\begin{itemize}
\item For given initial planet semi-major axis $a_0$ and binary
  orbital parameters ($a_{\rm out}$ and $e_{\rm out}$), calculate the
  ``migration eccentricity'' $e\mig=1-r_{p,{\rm mig}}/a_0$, with
  $r_{p,{\rm mig}}$ (the required pericenter distance in order for
  tidal dissipation to induce efficient orbital decay) given by
  \equ{eq:mig_distance}.

\item Knowing $e\mig$ and the octupole parameter $\epsilon_{\rm oct}$,
  calculate the critical planet-binary inclination angle for
  migration, $i_{0,\mathrm{mig}}$, using \equ{eq:solid_angle2}.
  
\item Assuming isotropic distribution of the binary orientations, the
  migration fraction is $f\mig^\mathrm{LK}=\cos i_{0,\mathrm{mig}}$.
  The tidal disruption fraction is similarly calculated from
  $f\dis^\mathrm{LK}=\cos i_{0,\mathrm{dis}}$ and the HJ (hot planet)
  formation fraction is
  $f\HJ^\mathrm{LK}=f\mig^\mathrm{LK}-f\dis^\mathrm{LK}$.

\item For a distribution of $a_0$, $a\out$ and $e\out$, the integrated
  migration/disruption fractions can be obtained using Eqs.~\ref{eq:integrated_fraction}
 and~\ref{eq:global_integrated_fraction}.
\end{itemize}

The results of our calculations are in good agreement with the recent
population studies of HJ formation via LK migration in stellar binaries
\citep{pet15a,and16}. They can be easily
modified to reflect different types of planets/binaries and
distributions of the binary's orbital elements.  This agreement
confirms that the total migration fraction of planets (the sum of
HJ and tidal disruption fractions) via the LK mechanism is
primarily determined by the initial geometry of the hierarchical triple,
and depends weakly on the properties of the planet \citep{and16}.
The production fraction of close-in planets, on the other hand, does depend
on the internal properties of the planet (e.g., mass-radius
relation and the strength of dissipation), and it decreases with
decreasing mass in the case of gas giants, resulting in a natural
absence of ``hot Saturns". In addition, we find that rocky planets,
provided they are sufficiently dissipative, may migrate more effectively than
Saturn-mass planets.
It is possible that some fraction of the observed short-period rocky planets
are produced via LK migration in stellar binaries.

\section*{acknowledgements}
We thank Kassandra Anderson for useful discussions and Crist\'obal
Petrovich for comments on the manuscript. This work has been
supported in part by NSF grant AST-1211061, and NASA grants NNX14AG94G
and NNX14AP31G.



\appendix


\begin{thebibliography}{43}
\expandafter\ifx\csname natexlab\endcsname\relax\def\natexlab#1{#1}\fi

\bibitem[{{Anderson}, {Storch} \& {Lai}(2016){Anderson}, {Storch}, \&
  {Lai}}]{and16}
{Anderson} K.~R., {Storch} N.~I., {Lai} D., 2016 (ASL), \mnras, 456, 3671

\bibitem[{{Beaug{\'e}} \& {Nesvorn{\'y}}(2012)}]{bea12}
{Beaug{\'e}} C., {Nesvorn{\'y}} D., 2012, \apj, 751, 119

\bibitem[{{Chatterjee} {et~al}\mbox{.}(2008){Chatterjee}, {Ford}, {Matsumura},
  \& {Rasio}}]{cha08}
{Chatterjee} S., {Ford} E.~B., {Matsumura} S., {Rasio} F.~A., 2008, \apj, 686,
  580

\bibitem[{{Correia} {et~al}\mbox{.}(2011){Correia}, {Laskar}, {Farago}, \&
  {Bou{\'e}}}]{cor11}
{Correia} A.~C.~M., {Laskar} J., {Farago} F., {Bou{\'e}} G., 2011, Celestial
  Mechanics and Dynamical Astronomy, 111, 105

\bibitem[{{Dawson} \& {Murray-Clay}(2013)}]{daw13}
{Dawson} R.~I., {Murray-Clay} R.~A., 2013, \apjl, 767, L24

\bibitem[{{Dawson}, {Murray-Clay} \& {Johnson}(2015){Dawson}, {Murray-Clay}, \&
  {Johnson}}]{daw15}
{Dawson} R.~I., {Murray-Clay} R.~A., {Johnson} J.~A., 2015, \apj, 798, 66

\bibitem[{{Fabrycky} \& {Tremaine}(2007)}]{fab07}
{Fabrycky} D., {Tremaine} S., 2007, \apj, 669, 1298

\bibitem[{{Ford}, {Kozinsky} \& {Rasio}(2000){Ford}, {Kozinsky}, \&
  {Rasio}}]{for00}
{Ford} E.~B., {Kozinsky} B., {Rasio} F.~A., 2000, \apj, 535, 385

\bibitem[{{Fressin} {et~al}\mbox{.}(2013){Fressin}, {Torres}, {Charbonneau},
  {Bryson}, {Christiansen}, {Dressing}, {Jenkins}, {Walkowicz}, \&
  {Batalha}}]{fre13}
{Fressin} F. {et~al.}, 2013, \apj, 766, 81

\bibitem[{{Goldreich} \& {Tremaine}(1980)}]{gol80}
{Goldreich} P., {Tremaine} S., 1980, ApJ, 241, 425

\bibitem[{{Guillochon}, {Ramirez-Ruiz} \& {Lin}(2011){Guillochon},
  {Ramirez-Ruiz}, \& {Lin}}]{gui11}
{Guillochon} J., {Ramirez-Ruiz} E., {Lin} D., 2011, \apj, 732, 74

\bibitem[{{Hamers}, {Perets} \& {Portegies Zwart}(2016){Hamers}, {Perets}, \&
  {Portegies Zwart}}]{ham16}
{Hamers} A.~S., {Perets} H.~B., {Portegies Zwart} S.~F., 2016, \mnras, 455,
  3180

\bibitem[{{Howard} {et~al}\mbox{.}(2012){Howard}, {Marcy}, {Bryson}, {Jenkins},
  {Rowe}, {Batalha}, {Borucki}, {Koch}, {Dunham}, {Gautier}, {Van Cleve},
  {Cochran}, \& {Latham}}]{how12}
{Howard} A.~W. {et~al.}, 2012, \apjs, 201, 15

\bibitem[{{Juri{\'c}} \& {Tremaine}(2008)}]{jur08}
{Juri{\'c}} M., {Tremaine} S., 2008, \apj, 686, 603

\bibitem[{{Katz}, {Dong} \& {Malhotra}(2011){Katz}, {Dong}, \&
  {Malhotra}}]{kat11}
{Katz} B., {Dong} S., {Malhotra} R., 2011, Physical Review Letters, 107, 181101

\bibitem[{{Kozai}(1962)}]{koz62}
{Kozai} Y., 1962, \aj, 67, 591

\bibitem[{{Lambeck}(1977)}]{lam77}
{Lambeck} K., 1977, Philosophical Transactions of the Royal Society of London
  Series A, 287, 545

\bibitem[{{Lidov}(1962)}]{lid62}
{Lidov} M.~L., 1962, P\&SS, 9, 719

\bibitem[{{Lin}, {Bodenheimer} \& {Richardson}(1996){Lin}, {Bodenheimer}, \&
  {Richardson}}]{lin96}
{Lin} D.~N.~C., {Bodenheimer} P., {Richardson} D.~C., 1996, \nat, 380, 606

\bibitem[{{Lithwick} \& {Naoz}(2011)}]{lit11}
{Lithwick} Y., {Naoz} S., 2011, \apj, 742, 94

\bibitem[{{Liu}, {Mu{\~n}oz} \& {Lai}(2015){Liu}, {Mu{\~n}oz}, \&
  {Lai}}]{liu15}
{Liu} B., {Mu{\~n}oz} D.~J., {Lai} D., 2015 (LML), \mnras, 447, 751

\bibitem[{{Marcy} {et~al}\mbox{.}(2005){Marcy}, {Butler}, {Fischer}, {Vogt},
  {Wright}, {Tinney}, \& {Jones}}]{mar05}
{Marcy} G., {Butler} R.~P., {Fischer} D., {Vogt} S., {Wright} J.~T., {Tinney}
  C.~G., {Jones} H.~R.~A., 2005, Progress of Theoretical Physics Supplement,
  158, 24

\bibitem[{{Martin}, {Mazeh} \& {Fabrycky}(2015){Martin}, {Mazeh}, \&
  {Fabrycky}}]{mar15}
{Martin} D.~V., {Mazeh} T., {Fabrycky} D.~C., 2015, \mnras, 453, 3554

\bibitem[{{Mu{\~n}oz} \& {Lai}(2015)}]{mun15b}
{Mu{\~n}oz} D.~J., {Lai} D., 2015, Proceedings of the National Academy of
  Science, 112, 9264

\bibitem[{{Nagasawa}, {Ida} \& {Bessho}(2008){Nagasawa}, {Ida}, \&
  {Bessho}}]{nag08}
{Nagasawa} M., {Ida} S., {Bessho} T., 2008, \apj, 678, 498

\bibitem[{{Naoz}, {Farr} \& {Rasio}(2012){Naoz}, {Farr}, \& {Rasio}}]{nao12}
{Naoz} S., {Farr} W.~M., {Rasio} F.~A., 2012, \apjl, 754, L36

\bibitem[{{Neron de Surgy} \& {Laskar}(1997)}]{ner97}
{Neron de Surgy} O., {Laskar} J., 1997, \aap, 318, 975

\bibitem[{{Peale}(1999)}]{pea99}
{Peale} S.~J., 1999, \araa, 37, 533

\bibitem[{{Petigura}, {Howard} \& {Marcy}(2013){Petigura}, {Howard}, \&
  {Marcy}}]{pet13}
{Petigura} E.~A., {Howard} A.~W., {Marcy} G.~W., 2013, Proceedings of the
  National Academy of Science, 110, 19273

\bibitem[{{Petrovich}(2015{\natexlab{a}})}]{pet15b}
{Petrovich} C., 2015{\natexlab{a}}, \apj, 805, 75

\bibitem[{{Petrovich}(2015{\natexlab{b}})}]{pet15a}
{Petrovich} C., 2015{\natexlab{b}}, \apj, 799, 27

\bibitem[{{Plavchan}, {Chen} \& {Pohl}(2015){Plavchan}, {Chen}, \&
  {Pohl}}]{pla15}
{Plavchan} P., {Chen} X., {Pohl} G., 2015, \apj, 805, 174

\bibitem[{{Raghavan} {et~al}\mbox{.}(2010){Raghavan}, {McAlister}, {Henry},
  {Latham}, {Marcy}, {Mason}, {Gies}, {White}, \& {ten Brummelaar}}]{rag10}
{Raghavan} D. {et~al.}, 2010, \apjs, 190, 1

\bibitem[{{Rasio} \& {Ford}(1996)}]{ras96}
{Rasio} F.~A., {Ford} E.~B., 1996, Science, 274, 954

\bibitem[{{Sanchis-Ojeda} {et~al}\mbox{.}(2014){Sanchis-Ojeda}, {Rappaport},
  {Winn}, {Kotson}, {Levine}, \& {El Mellah}}]{san14}
{Sanchis-Ojeda} R., {Rappaport} S., {Winn} J.~N., {Kotson} M.~C., {Levine} A.,
  {El Mellah} I., 2014, \apj, 787, 47

\bibitem[{{Seager} {et~al}\mbox{.}(2007){Seager}, {Kuchner}, {Hier-Majumder},
  \& {Militzer}}]{sea07}
{Seager} S., {Kuchner} M., {Hier-Majumder} C.~A., {Militzer} B., 2007, \apj,
  669, 1279

\bibitem[{{Steffen} \& {Farr}(2013)}]{ste13}
{Steffen} J.~H., {Farr} W.~M., 2013, \apjl, 774, L12

\bibitem[{{Storch}, {Anderson} \& {Lai}(2014){Storch}, {Anderson}, \&
  {Lai}}]{sto14}
{Storch} N.~I., {Anderson} K.~R., {Lai} D., 2014, {\it Science}, 345, 1317,
  preprint(arXiv:1409.3247)

\bibitem[{{Wright} {et~al}\mbox{.}(2012){Wright}, {Marcy}, {Howard}, {Johnson},
  {Morton}, \& {Fischer}}]{wri12}
{Wright} J.~T., {Marcy} G.~W., {Howard} A.~W., {Johnson} J.~A., {Morton} T.~D.,
  {Fischer} D.~A., 2012, \apj, 753, 160

\bibitem[{{Wu} \& {Lithwick}(2011)}]{wu11}
{Wu} Y., {Lithwick} Y., 2011, \apj, 735, 109

\bibitem[{{Wu} \& {Murray}(2003)}]{wu03}
{Wu} Y., {Murray} N., 2003, \apj, 589, 605

\bibitem[{{Wu}, {Murray} \& {Ramsahai}(2007){Wu}, {Murray}, \&
  {Ramsahai}}]{wu07}
{Wu} Y., {Murray} N.~W., {Ramsahai} J.~M., 2007, \apj, 670, 820

\bibitem[{{Yoder}(1995)}]{yod95}
{Yoder} C.~F., 1995, in Global Earth Physics: A Handbook of Physical Constants,
  {Ahrens} T.~J., ed., p.~1

\end{thebibliography}
 \end{document}